\documentclass[journal,12pt,onecolumn,draftclsnofoot,]{IEEEtran}
\usepackage{color}
\usepackage{graphicx}
\usepackage{amssymb}
\usepackage{amsmath}
\usepackage{amsthm}
\usepackage{etoolbox}
\usepackage{caption}
\usepackage{subcaption}
\usepackage[refpage]{nomencl}

\usepackage{mathtools}

\usepackage{blindtext}
\usepackage{enumitem}

\begin{document}
\title{QoE-Aware Beamforming Design for Massive MIMO Heterogeneous Networks \\ }
\author{Hadis Abarghouyi$^\ast$, S. Mohammad Razavizadeh$^\ast$, and Emil Bj\"{o}rnson$^\dag$ \ \\ \ \\
$^\ast$School of Electrical Engineering, Iran University of Science and Technology (IUST), 
Tehran, Iran \\
E-mail: hadis.abarghouyi@gmail.com, smrazavi@iust.ac.ir\\
$^\dag$ Department of Electrical Engineering (ISY), Link\"{o}ping University, Link\"{o}ping, Sweden\\
E-mail: emil.bjornson@liu.se
}
\maketitle
\color{black}
\begin{abstract}
One of the main goals of the future wireless networks is improving the users' quality of experience (QoE).
In this paper, we consider the problem of QoE-based resource allocation in the downlink of a massive multiple-input multiple-output (MIMO) heterogeneous network (HetNet). The network consists of a macro cell with a number of small cells embedded in it. 
The small cells' base stations (BSs) are equipped with a few antennas, while the macro BS is equipped with a massive number of antennas. We consider the two services Video and Web Browsing and design the beamforming vectors at the BSs. The objective is to maximize the aggregated Mean Opinion Score (MOS) of the users under constraints on the BSs' powers and the required quality of service (QoS) of the users. We also consider extra constraints on the QoE of users to more strongly enforce the QoE in the beamforming design. To reduce the complexity of the optimization problem, we suggest suboptimal and computationally efficient solutions. Our results illustrate that increasing the number of antennas at the BSs and also increasing the number of small cells' antennas in the network leads to a higher user satisfaction.
\end{abstract}

\section{Introduction}\label{sec:Introduction}
\IEEEPARstart{I}{n} future wireless networks (or 5G networks) the users will request more data traffic and diverse services than today.
Two important technologies that have been proposed for future wireless networks are small cells and massive multiple-input multiple-output (MaMIMO). Small cells can be used in combination with macro cells to form multi-tier or heterogeneous networks (HetNets) that can provide higher capacity and quality than conventional homogeneous networks \cite{r3}. On the other hand, MaMIMO is a technology in which base stations (BSs) in a cellular network are equipped with a large number of antennas (up to a few hundred) that can simultaneously serve a large number of users. MaMIMO can also be deployed in the HetNets for achieving higher performance, which is the case that we consider in this paper.

Service and network providers have studied various QoS metrics to optimize and enhance their network's performance. QoS metrics such as packet loss rate, transfer delay, throughput and coverage are mainly based on technical performance rather than users' experience. Recent studies show that although the conventional technical criteria based on the QoS are important, they are not sufficient for measuring the users' experience. In fact, the users' perception is affected by both technical and nontechnical (human-based) parameters \cite{r1-new}. Hence, for assuring better user experience, service providers have been switching their focus to perceived end to end quality, referred as Quality of Experience (QoE). ITU-T describes QoE as ``The overall acceptability of an application or service, as perceived subjectively by the end-user" \cite{TVT-traffic}.  

In general, QoE can be evaluated by subjective as well as objective methods. The subjective methods are based on evaluations given by human feelings about a service. One of the common subjective assessment methods is based on Mean Opinion Score (MOS) \cite{TVT-qoe-aware}. This parameter is a real number ranging from 1 to 5 which are related to bad, poor, fair, good and excellent, respectively.
Despite their advantages, the subjective assessments of QoE are usually time-consuming, difficult, costly and not real-time \cite{TVT-adhoc}. Therefore researchers have recently provided some objective QoE assessments models which are extracted from the QoS parameters. The objective assessment of QoE encompasses communication process measurements and could help service providers to indirectly estimate the users' satisfaction from technical parameters. The equations that relate the QoE to the QoS parameters are obtained by experimental measurements and mathematical analyses \cite{r8-new}-\cite{r16}. For example, in \cite{r8-new} the authors use the results of experimental measurements to establish some models such as linear, exponential and logarithmic functions. In \cite{r9-new}, the authors present a MOS model to map the page response time to QoE metrics using a Lorentzian function. In \cite{r16}, the authors propose models to describe the user perception of web browsing and video service in terms of some QoS parameters.

In wireless networks and especially in MaMIMO HetNets, the conventional criteria to allocate resources to users or improving the network performance  are based on the QoS parameters \cite{r12}-\cite{r14}.
However, recently researchers have begun to examine QoE concept for optimizing the network parameters. 
For example, in \cite{TWC1} the authors show that the QoE-based resource allocation is more efficient than the QoS-based methods in responding to users' demand \cite{relay}-\cite{Twc2}.
In \cite{relay}, cooperative networks have been studied and the QoE metric is used in optimizing the relay deployment in the network. In \cite{int16}, the authors propose a QoE based power allocation method for video streaming over wireless networks.
In \cite{r17}, the authors optimize the aggregated MOS of the users in a heterogeneous network consisted of one femtocell and one macro-cell. \cite{int23} proposes a beamforming method to maximize the aggregated MOS in cognitive radio networks. In \cite{Twc2}, the authors investigate the QoE improvement in a MIMO cognitive network. However, to the best of our knowledge, the QoE criterion has not been taken into account to improve the performance of the MaMIMO HetNets. 

In this paper, we propose a QoE-based joint beamforming and power allocation scheme for the downlink of a MaMIMO HetNet. The network is providing a web browsing or video streaming service to its users. These two services are expected to be among the basic services of the future 5G networks. Therefore it is important to provide an appropriate level of user satisfaction for these services \cite{r8-new}-\cite{r9-new}, \cite{hetero}. 
The heterogeneous network that we consider includes one macro cell with multiple small cells embedded in its coverage area. The macro base station (MBS) is equipped with an array with a large number of antennas. There are also multiple antennas at the small cell base stations (SBSs). This network serves a number of single antenna users.
We try to maximize the QoE of all users by optimizing the beamforming vectors and provide an efficient algorithm to solve the optimization problem. The objective of the optimization problem is to maximize the aggregated MOS of all users in the network, while satisfying constraints on the users' QoS and the power of the MBS and the SBSs. In addition, we include other constraints on the minimum QoE of the users to guarantee a minimum user satisfaction.

Our simulation results show that installing small cells in the network (i.e., using HetNet) leads to a higher users' satisfaction. On the other hand, increasing the number of antennas at the MBS has similar effects on the QoE. These results can be interesting for the network operators who seek solutions to fulfill their user satisfaction. In summary, the main contributions of this paper are as follow:

\begin{itemize}
  \item QoE-based Beamforming design for heterogeneous MaMIMO networks.
  \item Improving QoE for two different services of web browsing and video application.
  \item Considering subjective QoE parameter (MOS) as the objective and constraint of the optimization problem.
  \item Proposing new suboptimal but efficient method to solve two hard and non-convex optimization problems.
  \item Presenting the effect of adding small cells to the network or using a large number of antennas at the MBSs on the users' satisfaction in future 5G networks.
\end{itemize}

The rest of this paper is organized as follows: 
In Section \ref{sec:SYSTEM MODEL}, we describe the system model. Then we formulate the optimization problems in Section \ref{Problem Formulation}, to maximize the aggregated MOS of the network for the two adopted services. In Section \ref{sec:Optimal Power Allocation}, we solve the non-convex optimization problems by converting them to equivalent convex problems and using iterative algorithms. Finally, in Section \ref{sec:Simulation}, we illustrate the results by numerical examples.

\emph{Notations}:  $(.)^H$ and $\|.\|$ denote the conjugate transpose and Euclidean norm, respectively. $\mathcal{CN}(\boldsymbol{0},\boldsymbol{\boldsymbol{R}})$ denotes a circularly-symmetric complex Gaussian  random vector with zero mean and covariance matrix $\mathit{\boldsymbol{R}}$.
\section{System Model}\label{sec:SYSTEM MODEL}
\begin{figure}
\centering
\includegraphics[width=0.5\textwidth]{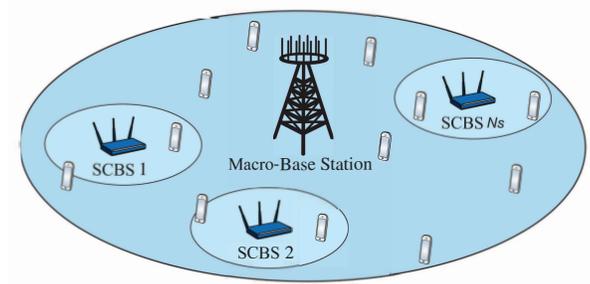}
\caption{Illustration of a MaMIMO HetNet consisted of one macro cell and $N_s$ small cells \cite{r14}.}
\label{Cell}
\end{figure}

In this paper, we consider the downlink of a two-tier MaMIMO HetNet consisting of $N_s$ small cells which are deployed in the coverage area of a macro cell as in Fig.\ref{Cell}. The MBS and SBSs use non-coherent joint transmission coordinated multipoint beamforming to provide a web browsing or video service for $K$ single antenna users. In this technique, the MBS and SBSs cooperate in transferring data to the users, but every BS sends a separate stream of data. The MBS is equipped with $M$ antennas and in this paper we are primarily interested in the MaMIMO regime where $M \gg K$ \cite{r14}. The number of antennas at the $j$th SBS is denoted by $N_j$.

As we mentioned earlier, MOS is a qualitative measure for assessing QoE, which can be expressed in terms of some objective parameters \cite{TVT-qoe-aware}. 
  For the adopted applications in this paper, an experimental relation between the QoE and QoS parameters can be expressed as follows. For web browsing, this relation can be expressed as \cite{r16}, 

\begin{equation}\label{web}
\mathrm{MOS^{web}}=-K_{1}\ln\left(d(R)\right)+K_{2}.
\end{equation}

In (\ref{web}), the constants $K_1$ and $K_2$ are selected in such a way that the value of $\mathrm{MOS^{web}}$ falls in the range of 1 to 5. In addition, $d(R)$[s] represents the page response time or the delay between a request for a web page and reception of the entire content of that web page.  $d(R)$ depends on parameters such as the web page size, round trip time (RTT) (the time interval that an IP packet travels from the server to the UE and returns \cite{r9-new}), and the type of utilized protocols (such as TCP and HTTP) which can be written as \cite{r16}
\begin{equation}\label{d(R)}
d(R)=3 \mathrm{RTT} + \frac{\mathrm{FS}}{B \cdot R} + L \left( \frac{\mathrm{MSS}}{B \cdot R}+\mathrm{RTT}\right) -\frac{2\mathrm{MSS}\left(2^L-1\right)}{B \cdot R},
\end{equation}
where $B$[Hz], $R$[bit/s/Hz], $\mathrm{FS}$[bit] and $\mathrm{MSS}$[bit] are the bandwidth, spectral efficiency, web page size and maximum segment size (the IP datagram size excluding the TCP/IP header \cite{r9-new}), respectively. $L=\min[L_1, L_2]$ is a parameter that specifies the number of slow start cycles with idle periods (the cycles in packet exchange between UE and server during web page download \cite{r9-new}), where $L_1$ and $L_2$ are defined as \cite{r16}
\begin{equation}\label{L1}
L_1=\log_2 \left( \frac{1}{2}+\frac{B \cdot R \cdot \mathrm{RTT}}{2\mathrm{MSS}}\right),
L_2=\log_2\left( \frac{1}{2}+\frac{\mathrm{FS}}{4\mathrm{MSS}}\right).
\end{equation}
For video service (H.264/MPEG-4 Video Coding), the MOS is obtained as \cite{videoconstants}

\begin{equation}\label{videoQoE}
\mathrm{MOS^{video}}=g\log\left(\mathrm{PSNR}\right)+e,
\end{equation}
\noindent where $g$ and $e$ are two constants  that are selected in such a way that the value of $\mathrm{MOS^{video}}$ falls in the range of 1 to 5. $\mathrm{PSNR}$ shows peak SNR and is defined as 
\begin{equation}\label{psnr}
\mathrm{PSNR}=u+v\sqrt{\frac{B\cdot R}{r}}\left(1-\frac{r}{B \cdot R}\right).
\end{equation}
\noindent $u$, $v$ and $r$ parameters also characterize a specific video stream.

All subchannels between the users and the MBS or SBSs are modeled as flat fading channels. The channel between the $k$th user and the MBS  and between the $k$th user and the $j$th SBS is denoted by $\mathbf{h}_{k,0}\in\mathbb{C}^{M\times 1}$ and $\mathbf{h}_{k,j}\in\mathbb{C}^{N_j\times 1}$, respectively. We assume that perfect channel state information is available at the BSs. The transmitted signal vectors from the MBS and the $j$th SBS are represented by $\mathbf{x}_0\in\mathbb{C}^{M\times 1}$ and $\mathbf{x}_j\in\mathbb{C}^{N_j\times 1}$, respectively. The received signal at the $k$th user is modeled as
\begin{equation}\label{yk}
{y}_{k}=\mathbf{h}_{k,0}^H \mathbf{x}_{0}+\sum_{j=1}^{N_s}\mathbf{h}_{k,j}^H \mathbf{x}_{j}+n_{k} ,
\end{equation}
where $n_k \sim \mathcal{CN}(0,\sigma_k^2(mW))$ is the Additive White Gaussian Noise (AWGN) at the receiver.

The transmitted signals $\mathbf{x}_0$ and $\mathbf{x}_j$ are obtained by applying appropriate beamforming vectors at the BSs as
\begin{equation}\label{x0}
\mathbf{x}_{j}=\sum_{l=1}^{K}\boldsymbol{w}_{l,j} d_{l,j}   \   \ ,  \ j=0.1,...,N_s,
\end{equation}
\noindent where $\boldsymbol{w}_{l,0}\in\mathbb{C}^{M\times1}$ and $\boldsymbol{w}_{l,j}\in\mathbb{C}^{N_j\times1} (j=1,...,N_s)$ denote the beamforming vectors at the MBS and the $j$th SBS corresponding to the $l$th user, respectively, $d_{l,j}$ is the information symbol transmitted to the $l$th user by the $j$th BS ($j=0$ is related to the MBS). It is assumed that the information symbols are independent and have unit power (1 mw). In the following sections, we show how to efficiently design the beamforming vectors and the transmitted power of the base stations.

\section{ Problem Formulation}\label{Problem Formulation}
In this section, to find the optimum beamforming based on QoE maximization, we consider two problems corresponding to two different adopted services. The target of the two problems is to maximize the aggregated MOS of the users subject to some constraints on QoS and QoE of users. 

Since the value of $\mathrm{RTT}$ in 5G networks is very small and we also consider only a few subcarriers, we can ignore it in calculating the MOS parameter in (\ref{web}) \cite{r16}. Hence, for web services the $\mathrm{MOS}$ of $k$th user can be represented by
\begin{equation}\label{ignorerttMOS}
\mathrm{MOS}_k^{web}(\mathtt{w})=K_{1}\ln\bigg( \frac{B \cdot R_k(\mathtt{w})}{\mathrm{FS}_k}\bigg)+K_{2},
\end{equation}

\noindent where $\mathtt{w}=\{\boldsymbol{w}_{k,j}\ | \ k=1,...,K, \ j=0,1,...,N_s\}$ is the set of beamforming vectors. For video service, we have

\begin{equation}\label{videowww}
 \mathrm{MOS_k^{video}}(\mathtt{w})=g\log\left(\mathrm{PSNR}(\mathtt{w})\right)+e,
\end{equation}
\noindent where
\begin{equation}\label{psnrrrr}
\mathrm{PSNR}(\mathtt{w})=u+v\sqrt{\frac{B\cdot R_k(\mathtt{w})}{r}}\left(1-\frac{r}{B \cdot R_k(\mathtt{w})}\right).
\end{equation}

We also assume constraints on the users' QoS which are defined in terms of the minimum required data rate of $R_{k,min}$
\begin{equation}
B\cdot R_k(\mathtt{w})  
\geq R_{k,min},  \ \forall k
\end{equation}
\noindent where
\begin{equation}\label{eqn7}
R_k(\mathtt{w}) =\log_2 \bigg(1+ \frac{\sum_{j=0}^{N_s}\mid\mathbf{h}_{k,j}^H \boldsymbol{w}_{k,j} \mid^2}{\sum_{j=0}^{N_s}\sum_{l=1,l\neq k}^{K}\mid\mathbf{h}_{k,j}^H \boldsymbol{w}_{l,j} \mid^2+\sigma_{k}^2} \bigg), 
\end{equation}
is the achievable sum spectral efficiency of the $k$th user, when the user is decoding data streams from the BSs in a sequential manner using successive interference cancelation \cite{r14}.

In this paper, we consider the per-antenna power constraints at all BSs which is more practical than total power constraints when each antenna has its own radio frequency chain \cite{perantenna}. The per-antenna power constraints for the MBS and the $j$th SBS can also be expressed, respectively, as
\begin{equation}\label{eqn5}
\sum_{k=1}^K\boldsymbol{w}_{k,0}^H\mathbf{D}_{q,0} \boldsymbol{w}_{k,0}  \leq P_{0,q}  \  q=1,...,M,
\end{equation}
\begin{equation}\label{eqn6}
\begin{aligned}
\sum_{k=1}^K\boldsymbol{w}_{k,j}^H\mathbf{D}_{q,j} \boldsymbol{w}_{k,j}  \leq P_{j,q}  \ , \ &  j=1,...,N_s\\
& q=1,...,N_j,
\end{aligned}
\end{equation}
\noindent where $\mathbf{D}_{q,0}\in\mathbb{C}^{M\times M}$ and $\mathbf{D}_{q,j}\in\mathbb{C}^{N_j\times N_j}$ are positive semidefinite zero weighting matrices with only one $'1'$ at the $q$th diagonal element. $P_{0,q} $ and $P_{j,q} $ ($P_{0,q} \gg P_{j,q} \ ,j=1, \ldots, N_s$) represent the maximum transmitted powers from the $q$th antenna of the MBS and the $j$th SBS, respectively.

Our objective is to maximize the aggregated MOS of all users. However maximizing a aggregated MOS does not guarantee that every user will be satisfied. Hence QoE constraints are added to the problems to more strongly enforce the QoE in the beamforming design and ensure fairness between the users. This is done by defining a minimum satisfaction threshold for each user as below
\begin{equation}\label{eqnminmosweb}
\mathrm{MOS}_k^{web}(\mathtt{w}) \geq \mathrm{MOS}^{web}_{k,min},
\end{equation}
\noindent and
\begin{equation}\label{eqnminmosvideo}
\mathrm{MOS}_k^{video}(\mathtt{w}) \geq \mathrm{MOS}^{video}_{k,min}.
\end{equation}
\noindent From now on, the MOS of $k$th user for both services are represented by $\mathrm{MOS}_k(\mathtt{w})$. Using (\ref{ignorerttMOS})-(\ref{eqnminmosvideo}), the optimization problem to determine the beamforming vectors is written as
 
\begin{align}
& \underset{\boldsymbol{w}_{k,j} \forall k,j}{\text{maximize}}
& & \sum_{k=1}^K \mathrm{MOS}_{k}(\mathtt{w})\label{majoreqn}
\\
& \text{s.t.} & &
\sum_{k=1}^K\boldsymbol{w}_{k,0}^H\mathbf{D}_{q,0} \boldsymbol{w}_{k,0}  \leq P_{0,q}  \  , \ q=1,...,M \tag{\ref{majoreqn}.a} \label{majoreqna} \ \\ 
& \ & & \sum_{k=1}^K\boldsymbol{w}_{k,j}^H\mathbf{D}_{q,j} \boldsymbol{w}_{k,j}  \leq P_{j,q}  \  , j=1,...,N_s\tag{\ref{majoreqn}.b} \label{majoreqnb}\\
& & & \qquad \qquad  \qquad \qquad  \qquad \quad \ \ q=1,...,N_j\nonumber\\
& & & B\cdot R_k(\mathtt{w}) \geq R_{k,min} \tag{\ref{majoreqn}.c} \label{majoreqnc}\\
& & &  MOS_k(\mathtt{w}) \geq MOS_{min,k}, \tag{\ref{majoreqn}.d} \label{majoreqnd}
\end{align}

\noindent The problem is non-convex and hence, it cannot be solved efficiently. In next section, we will present a method for converting it to a convex optimization problem.
\section{Optimal Beamforming and Power Allocation }\label{sec:Optimal Power Allocation}
In this section, we present methods for solving the optimization problem in (\ref{majoreqn}) and therefore designing the beamforming vectors and the transmitted power of the MBS and SBSs. 
\subsection{Web Browsing Service}\label{web algorithm}
By considering $R_{k,min}=B\cdot \log_2\left(1+{\mathrm{SINR}_{k,min}}\right)$ and simple manipulation of the QoS constraints, the optimization problem in (\ref{majoreqn}) is converted to 
\begin{align}\label{secondproblem2}
& \underset{\boldsymbol{w_{k,j}} \forall k,j}{\text{maximize}}
& & \sum_{k=1}^K K_{1}\ln\bigg(\frac{B \cdot R_{k}(\mathtt{w})}{\mathrm{FS}_k} \bigg)+K_{2} \\
 & \text{s.t.} & & (\ref{majoreqna})-(\ref{majoreqnb}) \nonumber\\
& & & \frac{\sum_{j=0}^{N_s}\mid\mathbf{h}_{k,j}^H \boldsymbol{w}_{k,j} \mid^2}{\sum_{j=0}^{N_s}\sum_{l=1,l\neq k}^{K}\mid\mathbf{h}_{k,j}^H \boldsymbol{w}_{l,j} \mid^2+\sigma_{k}^2}
 \geq \mathrm{SINR}_{k,min} \nonumber\\
& & & \frac{\sum_{j=0}^{N_s}\mid\mathbf{h}_{k,j}^H \boldsymbol{w}_{k,j} \mid^2}{\sum_{j=0}^{N_s}\sum_{l=1,l\neq k}^{K}\mid\mathbf{h}_{k,j}^H \boldsymbol{w}_{l,j} \mid^2+\sigma_{k}^2}
\geq A_k,\nonumber
\end{align}

\noindent
 where $A_k=2^{\frac{\mathrm{FS}_k}{B}\cdot \exp\left(\frac{\mathrm{MOS}_{min,k}-K_2}{K_1}\right)}-1$.
\noindent Since the objective function in (\ref{secondproblem2}) is not concave and the two last constraints also are not convex, the problem is a non-convex problem. The constraints can be converted to convex constraints by defining some positive semidefinite matrices as 
\begin{equation}\label{Wkj}
  \mathbf{W}_{k,j}=\boldsymbol{w}_{k,j}\boldsymbol{w}_{k,j}^H,
\end{equation}
\noindent where $\mathrm{rank}(\mathbf{W}_{k,j})\leq1$\footnote{Note that $\mathrm{rank}(\mathbf{W}_{k,j})=0$ implies $\mathbf{W}_{k,j}=0$.}. These constraints lead to unique $\boldsymbol{w}_{k,j}$ in above equation (Lemma 3 in \cite{rankw}). Hence, we have

\begin{align}\label{eqn11}
& \underset{\boldsymbol{W}_{k,j} \forall k,j}{\text{maximize}}
& & \sum_{k=1}^K  K_{1}\ln\bigg(\frac{B \cdot R_k(\mathtt{W})}{\mathrm{FS}_k}\bigg)+K_{2} \\
& \text{s.t.}& & \sum_{k=1}^K \mathrm{tr}\left(\mathbf{D}_{q,0}\mathbf{W}_{k,0}\right) \leq P_{0,q}  \  q=1,...,M  \tag{\ref{eqn11}.a} \label{eqn11a} \\ \quad 
&  & & \sum_{k=1}^K \mathrm{tr}\left(\mathbf{D}_{q,j}\mathbf{W}_{k,j}\right) \leq P_{j,q} \ j=1,...,N_s  \nonumber\\ 
& & & \qquad \qquad  \qquad \qquad \quad  \ q=1,...,N_j \tag{\ref{eqn11}.b}\label{eqn11b}   \nonumber\\
& & &\bigg(\sum_{j=0}^{N_s} \mathbf{h}_{k,j}^H \left(1+\frac{1}{\mathrm{SINR}_{k,min}}\right)\mathbf{W}_{k,j}\mathbf{h}_{k,j}\nonumber \\ & & &  \ -\sum_{j=0}^{N_s} \sum_{l=1}^K \mathbf{h}_{k,j}^H \mathbf{W}_{l,j}\mathbf{h}_{k,j}\bigg) \geq \sigma_{k}^2 \tag{\ref{eqn11}.c} \label{eqn11c} \nonumber\\
& & &\bigg(\sum_{j=0}^{N_s} \mathbf{h}_{k,j}^H \left(1+\frac{1}{A_k} \right)\mathbf{W}_{k,j}\mathbf{h}_{k,j}\nonumber\\ & & &  \ -\sum_{j=0}^{N_s} \sum_{l=1}^K \mathbf{h}_{k,j}^H \mathbf{W}_{l,j}\mathbf{h}_{k,j}\bigg) \geq \sigma_{k}^2 \tag{\ref{eqn11}.d}\label{eqn11d}\\
& & & \mathrm{rank}(\mathbf{W}_{k,j})\leq 1  \tag{\ref{eqn11}.e}
\label{eqn11e},
\end{align}
\noindent where
\begin{equation}\label{ratew}
 R_k(\mathtt{W})=\\
\log_2\left( 1+\frac{\sum_{j=0}^{N_s}\mathbf{h}_{k,j}^H \boldsymbol{W}_{k,j} \mathbf{h}_{k,j}}{\sum_{j=0}^{N_s}\sum_{l=1,l\neq k}^{K}\mathbf{h}_{k,j}^H \boldsymbol{W}_{l,j}\mathbf{h}_{k,j} +\sigma_{k}^2}\right),
\end{equation}

\noindent and  $\mathtt{W}=\{\boldsymbol{W}_{k,j}\ | \ k=1,...,K, \ j=0,1,...,S\}$ is the set of beamforming matrices.
The rank constraints and the objective function are not convex and concave, respectively. Therefore, we consider an equivalent convex optimization problem by calculating the superlevel sets of the objective function as
 \begin{equation}\label{eqnobjective2}
\begin{aligned}
& \underset{\boldsymbol{W}_{k,j}, z_k \forall k,j}{\text{maximize}}
& & \sum_{k=1}^K K_{1}\ln(z_k )+K_{2} \\
& \text{s.t.} & &  \frac{B \cdot R_k(\mathtt{W})}{\mathrm{FS}_k} \geq z_k
,\quad z_k >0  \\
& & & (\ref{eqn11a})-(\ref{eqn11e}).
\end{aligned}
\end{equation}
The new objective function is concave, but the new constraints are still non-convex. We convert them to convex functions by introducing additional optimization variables.

Defining the lower bound for the spectral efficiency of each user $R_k(\mathtt{W}) $ as $\log_2 (t_k)$, (\ref{eqnobjective2}) can be rewritten as

\begin{align}\label{eqnobjective3}
& \underset{\boldsymbol{W}_{k,j},z_k,t_k \forall k,j}{\text{maximize}}
& & \sum_{k=1}^K K_{1}\ln(z_k )+K_{2} \\
& \text{s.t.} & &\log_2(t_k) \geq \frac{z_k\cdot \mathrm{FS}_k}{B}  \tag{\ref{eqnobjective3}.a} \label{eqnobjective3a}\\
& & & \frac{\sum_{j=0}^{N_s}\mathbf{h}_{k,j}^H \boldsymbol{W}_{k,j} \mathbf{h}_{k,j}}{\sum_{j=0}^{N_s}\sum_{l=1,l\neq k}^{K}\mathbf{h}_{k,j}^H \boldsymbol{W}_{l,j}\mathbf{h}_{k,j} +\sigma_{k}^2} \geq t_k-1  \tag{\ref{eqnobjective3}.b} \label{eqnobjective3b}\\
 & & & t_k >0 , \quad  z_k >0    \tag{\ref{eqnobjective3}.c} \label{eqnobjective3c}\\
& & & (\ref{eqn11a})-(\ref{eqn11e}). \tag{\ref{eqnobjective3}.d} \label{eqnobjective3d}
\end{align}

\noindent Considering that $\sum_{j=0}^{N_s}\sum_{l=1,l\neq k}^{K}\mathbf{h}_{k,j}^H \boldsymbol{W}_{l,j}\mathbf{h}_{k,j} +\sigma_{k}^2 \leq s_k,$ (\ref{eqnobjective3}) can be converted to the following
\begin{align}\label{eqnobjective4}
& \underset{\boldsymbol{W}_{k,j},z_k,t_k,s_k \forall k,j}{\text{maximize}}
& & \sum_{k=1}^K K_{1}\ln(z_k )+K_{2}  \\
& \text{s.t.} & & \log_2(t_k) \geq \frac{z_k\cdot \mathrm{FS}_k}{B}   \tag{\ref{eqnobjective4}.a} \label{eqnobjective4a}\\
& & & \sum_{j=0}^{N_s} \mathbf{h}_{k,j}^H \mathbf{W}_{k,j}\mathbf{h}_{k,j} \geq (t_k-1)  s_k \tag{\ref{eqnobjective4}.b} \label{eqnobjective4b}  \\
& & & \sum_{j=0}^{N_s} \sum_{l=1,l\neq k}^K \mathbf{h}_{k,j}^H\mathbf{W}_{l,j}\mathbf{h}_{k,j}+\sigma_{k}^2\leq s_k \tag{\ref{eqnobjective4}.c} \label{eqnobjective4c} \\
 & & & (\ref{eqnobjective3c})-(\ref{eqnobjective3d}).\tag{\ref{eqnobjective4}.d} \label{eqnobjective4d}
\end{align}
  \noindent This problem is still non-convex, because of the quasiconcave form of the $t_k s_k$ function. Therefore, we replace it by a looser upper bound which is a convex function. For any $\lambda_k >0$ the bound is defined as \cite{r19}
  
  \begin{equation}\label{implibound}
\frac{\lambda_k}{2}t_k^2+ \frac{1}{2\lambda_k}s_k^2 \geq t_k s_k,
 \end{equation}
 \noindent where the equality is satisfied by $\lambda_k = \frac{s_k}{t_k}$.

By using (\ref{secondproblem2})--(\ref{implibound}), the optimization problem in (\ref{eqnobjective4}) can be written as equation (\ref{eqn12}).
  \begin{align}\label{eqn12}
& \underset{\boldsymbol{W}_{k,j},z_k,t_k,s_k \forall k,j}{\text{maximize}}
& & \sum_{k=1}^K K_{1}\ln(z_k )+K_{2}\\
& \text{s.t.} & & \log_2(t_k)  \geq \frac{z_k\cdot \mathrm{FS}_k}{B}  \tag{\ref{eqn12}.a} \label{eqn12a} \\
& & & \sum_{j=0}^{N_s} \mathbf{h}_{k,j}^H \mathbf{W}_{k,j}\mathbf{h}_{k,j} \geq \frac{\lambda_k}{2}t_k^2+ \frac{1}{2\lambda_k}s_k^2 -s_k \tag{\ref{eqn12}.b} \label{eqn12b}\\
& & & (\ref{eqnobjective4c})-(\ref{eqnobjective4d}). \tag{\ref{eqn12}.c} \label{eqn12c}\
 \end{align}
 By relaxing the constraint $\mathrm{rank}(\mathbf{W}_{k,j})\leq1$, the optimization problem is translated to (\ref{relaxed-web}), whose solutions are the same as the original problem. It is proved in Lemma 3 \cite{rankw} that the solution of this problem will surly satisfy the rank condition. 

 \begin{align}\label{relaxed-web}
& \underset{\boldsymbol{W}_{k,j},z_k,t_k,s_k \forall k,j}{\text{maximize}}
& & \sum_{k=1}^K K_{1}\ln(z_k )+K_{2}\\
& \text{s.t.} & & (\ref{eqn12a})-(\ref{eqn12b})\nonumber\\
& & & (\ref{eqnobjective4c}) \nonumber\\
& & & (\ref{eqnobjective3c}) \nonumber\\
 & & & (\ref{eqn11a})-(\ref{eqn11d}).\nonumber
 \end{align}
All the constraint in (\ref{relaxed-web}) are convex, and hence, this problem can be solved efficiently by standard techniques. Because of $\lambda_k$ in (\ref{implibound}), there may be a difference between the optimal solutions of (\ref{majoreqn}) and (\ref{relaxed-web}). In Table~\ref{table}, we propose an iterative algorithm to reduce this. It should be noted that our proposed algorithm is a Sequential Parametric Convex Approximation as described in \cite{converge}. In addition, based on Proposition 3.2 in \cite{converge}, a KKT point to the problem is achieved. 
\small
\begin{table}
\caption{\label{table}}
\begin{center}
\begin{tabular}{|p{8cm}|}
\hline
Algorithm 1: The proposed algorithm for solving equation (\ref{majoreqn})\\
\\
\hline 
1. Initialize $\lambda_k > 0 , \forall k$  such that  the optimization variables belong to the feasible set of (\ref{relaxed-web}).\\
2. Solve (\ref{relaxed-web}) to obtain the optimal solutions of $t_k$ and $s_k$ $\forall k$ (called $t_k^*$ and $s_k^*$ ).\\
3.Update $\lambda_k$ by using the $\lambda_k^*=\frac{s_k^*}{t_k^*}  , \forall k$.\\
4.If $|\lambda_k^* - \lambda_k|\leq \epsilon $, where $\epsilon$ is a predefined threshold, stop the algorithm. Else replace the $\lambda_k$ by $\lambda_k^*$ and return to step 2.\\
\hline
\end{tabular}
\end{center}
\end{table}
\normalsize

\subsection{Video Service}\label{video_algorithm}
For video service, we use (\ref{videowww}) to write the optimization problem in (\ref{majoreqn}) as follows:
\begin{align}
& \underset{\boldsymbol{w}_{k,j} \forall k,j}{\text{maximize}}
& & \sum_{k=1}^K \mathrm{MOS}_{k}(\mathtt{w})\label{videow}
\\
& \text{s.t.} & &
\sum_{k=1}^K\boldsymbol{w}_{k,0}^H\mathbf{D}_{q,0} \boldsymbol{w}_{k,0}  \leq P_{0,q}  \  , \ q=1,...,M \tag{\ref{videow}.a} \label{videowa} \ \\ 
& \ & & \sum_{k=1}^K\boldsymbol{w}_{k,j}^H\mathbf{D}_{q,j} \boldsymbol{w}_{k,j}  \leq P_{j,q}  \  , j=1,...,N_s\tag{\ref{videow}.b} \label{videowb}\\
& & & \qquad \qquad  \qquad \qquad  \qquad \quad \ \ q=1,...,N_j\nonumber\\
 & & & \frac{\sum_{j=0}^{N_s}\mid\mathbf{h}_{k,j}^H \boldsymbol{w}_{k,j} \mid^2}{\sum_{j=0}^{N_s}\sum_{l=1,l\neq k}^{K}\mid\mathbf{h}_{k,j}^H \boldsymbol{w}_{l,j} \mid^2+\sigma_{k}^2}
 \geq \mathrm{SINR}_{k,min} \tag{\ref{videow}.c} \label{videowc}\\
& & & \frac{\sum_{j=0}^{N_s}\mid\mathbf{h}_{k,j}^H \boldsymbol{w}_{k,j} \mid^2}{\sum_{j=0}^{N_s}\sum_{l=1,l\neq k}^{K}\mid\mathbf{h}_{k,j}^H \boldsymbol{w}_{l,j} \mid^2+\sigma_{k}^2}
\geq A_k \tag{\ref{videow}.d} \label{videowd}
\end{align}
\noindent
where $A_k=2^{\left(\frac{\frac{\sqrt{B\cdot r}}{v} X_k+\sqrt{\frac{B\cdot rX_k^2}{v^2}+4r\cdot B}}{2B}\right)^2}-1$ is obtained as: 
 \begin{align}\label{minMOSv}
 & g\log\left(u+v\sqrt{\frac{B\cdot R_k(\mathtt{W})}{r}}\left(1-\frac{r}{B \cdot R_k(\mathtt{W}) }\right)\right)+e \geq\nonumber \\
 & \qquad \qquad \qquad  \qquad \qquad \qquad \qquad \qquad \qquad \mathrm{MOS}_{k,min}
\end{align}

\noindent By simplifying this equation,

\begin{equation}
 B\cdot R_k(\mathtt{W})- \frac{\sqrt{B\cdot R_k(\mathtt{W}) \cdot r}}{v} X_k-r \geq 0
\end{equation}
\noindent
where $X_k=10^{\left(\frac{\mathrm{MOS}_{k,min}-e}{g}\right)}-u.$ The $A_k$ is obtained by solving this equation and using (\ref{ratew}).
 
Then from (\ref{Wkj}), the problem in (\ref{videow}) is converted to
\begin{align}\label{videow2}
& \underset{\boldsymbol{W}_{k,j} \forall k,j}{\text{maximize}}
& & \sum_{k=1}^K  \mathrm{MOS}_{k}(\mathtt{W}) \\
& \text{s.t.}& & (\ref{eqn11a})-(\ref{eqn11e}) \nonumber
\end{align}
\noindent
where $\mathtt{W}=\{\boldsymbol{W}_{k,j}\ | \ k=1,...,K, \ j=0,1,...,S\}$ is the set of beamforming matrices and

\begin{equation}\label{sumMOSW}
\begin{aligned}
&\mathrm{MOS}_{k}(\mathtt{W})=\\
& \left(g\log\left(u+v\sqrt{\frac{B\cdot R_k(\mathtt{W}) }{r}}(1-\frac{r}{B \cdot R_k(\mathtt{W}) })\right)+e\right) 
\end{aligned}
\end{equation}
\noindent
where $R_k(\mathtt{W})$ is defined in (\ref{ratew}).

The rank constraints and the objective function are not convex and concave, respectively. Therefore, we consider an equivalent convex optimization problem by calculating the superlevel sets of the objective function.
Assume
\begin{align}
& u+v\sqrt{\frac{B\cdot R_{k}(\mathtt{W})}{r}}\left(1-\frac{r}{B\cdot R_{k}(\mathtt{W})}\right) \geq z_k \label{z2k}
\end{align}
\noindent and also
\begin{align} 
&\sqrt{B \cdot R_k(\mathtt{W})} \geq t_{1k} \label{t1k}\\
&R_k(\mathtt{W})\geq \log_2(t_{2k})  \tag{\ref{t1k}.a} \label{t1ka}\\
&t_{2k} >0.  \tag{\ref{t1k}.b} \label{t1kb}
\end{align}
\noindent
Then $B \cdot \log_2(t_{2k}) \geq t_{1k}^2$
 and (\ref{z2k}) is converted to a convex equation as $ t_{1k}-\frac{r}{t_{1k}}  \geq \frac{z_{k}-u}{v} \sqrt{r}.$
\noindent By using (\ref{t1ka}) and (\ref{ratew}), 
\begin{align} \label{RWvideo}
&\frac{\sum_{j=0}^{N_s}\mathbf{h}_{k,j}^H \boldsymbol{W}_{k,j} \mathbf{h}_{k,j}}{\sum_{j=0}^{N_s}\sum_{l=1,l\neq k}^{K}\mathbf{h}_{k,j}^H \boldsymbol{W}_{l,j}\mathbf{h}_{k,j} +\sigma_{k}^2} \geq t_{2k}-1.
\end{align}
\noindent 
By considering $\sum_{j=0}^{N_s}\sum_{l=1,l\neq k}^{K}\mathbf{h}_{k,j}^H \boldsymbol{W}_{l,j}\mathbf{h}_{k,j} +\sigma_{k}^2 \leq s_{k},$ (\ref{RWvideo}) is converted to
\begin{align} \label{t3s2ka}
 &\sum_{j=0}^{N_s}\mathbf{h}_{k,j}^H \boldsymbol{W}_{k,j} \mathbf{h}_{k,j} \geq s_{k}(t_{2k} -1).
\end{align}
\noindent
Using (\ref{implibound}), (\ref{t3s2ka}) is converted to the convex constraints
\begin{align} \label{landa2}
\begin{split}
& \sum_{j=0}^{N_s}\mathbf{h}_{k,j}^H \boldsymbol{W}_{k,j} \mathbf{h}_{k,j} \geq \frac{\lambda_{k}}{2}t_{2k}^2+ \frac{1}{2\lambda_{k}}s_{k}^2-s_{k}.
\end{split}
\end{align}
Therefore, by using (\ref{z2k})-(\ref{landa2}), (\ref{videow2}) is converted to
  \begin{align}
& \underset{\boldsymbol{W}_{k,j},z_k,t_{1k},t_{2k},s_k \forall k,j}{\text{maximize}}
& & \sum_{k=1}^K \left(g\log(z_k )+e\right) \label{videocvx1}\\
& \text{s.t.} & & t_{1k}-\frac{r}{t_{1k}}  \geq \frac{z_{k}-u}{v} \sqrt{r}  \tag{\ref{videocvx1}.a} \label{videocvx1a} \\
& & & B \cdot \log_2(t_{2k}) \geq t_{1k}^2  \tag{\ref{videocvx1}.b} \label{videocvx1b} \\
& & & \sum_{j=0}^{N_s} \mathbf{h}_{k,j}^H \mathbf{W}_{k,j}\mathbf{h}_{k,j} \geq  \nonumber \\
& & & \quad  \{\frac{\lambda_k}{2}t_{2k}^2+ \frac{1}{2\lambda_k}s_k^2-s_k\} \tag{\ref{videocvx1}.c} \label{videocvx1c}\\
& & & \sum_{j=0}^{N_s}\sum_{l=1,l\neq k}^{K}\mathbf{h}_{k,j}^H \boldsymbol{W}_{l,j}\mathbf{h}_{k,j} +\sigma_{k}^2 \leq s_{k} \tag{\ref{videocvx1}.d} \label{videocvx1d} \\
 & & &t_{1k} >0 , \quad t_{2k} >0 , \quad  z_k >0  \tag{\ref{videocvx1}.e} \label{videocvx1e}\\
   & & & (\ref{eqn11a})-(\ref{eqn11e}).\tag{\ref{videocvx1}.f} \label{videocvx1f}
 \end{align}
 By removing the constraint  (\ref{eqn11e}), the problem (\ref{videocvx1}) is converted to a relaxed problem as 

  \begin{align}
& \underset{\boldsymbol{W}_{k,j},z_k,t_{1k},t_{2k},s_k \forall k,j}{\text{maximize}}
& & \sum_{k=1}^K \left(g\log(z_k )+e\right) \label{videorelax}\\
& \text{s.t.} & & (\ref{videocvx1a})- (\ref{videocvx1e})\nonumber\\
 & & & (\ref{eqn11a})-(\ref{eqn11d}).\nonumber
 \end{align}
All the constraint functions of (\ref{videocvx1}) are convex and hence the problem can be solved efficiently by standard techniques. In  table \ref{table}. the proposed algorithm for solving this problem is shown which replaces $t_{k}$, $t_{k}^*$ and (\ref{relaxed-web}) with $t_{2k}$, $t_{2k}^*$ and (\ref{videorelax}).
The simulation results of proposed algorithms for web browsing and video will be given in the next section.

\section{Numerical Results}\label{sec:Simulation}
This section numerically evaluates the proposed schemes in the previous sections. We consider the downlink of a HetNet consisting of one macro cell with radius $500$ [m] and four small cells each with radius $40$ [m] that are deployed at the same area (see Fig. \ref{Cell}). The four SBSs are equally spaced on a circle of radius $250$ [m], centered at the MBS. We assume that there are 6 users in the Macro cell and one user in each small cell (total users $K=10$). The users are uniformly distributed in the coverage area (between the radiuses of $35$ [m] and $500$ [m] for Macro cell users and between the radiuses of $3$ [m] and $40$ [m] for small cell users). We assume that all the SBSs have equal number of antennas, i.e. $N_j=N,$ for $j=1,...,N_s$ where $N=1,2,3$. The maximum power at all antennas at the SBSs or MBS are $-10.9$ [dBm] and $18$ [dBm], respectively. Assume that the bandwidth of all subcarriers is $15$ [kHz] \cite{r14}. The path and penetration loss at a distance $d$ [km] from the MBS and  SBSs are $148.1+37.6\log_{10} (d)$ [dB] and $127+30\log_{10} (d) $ [dB], respectively. We consider standard deviation of $7$ [dB] for log-normal shadow fading \cite{r14}. We model the small scale fading channels by independent Rayleigh variables as $\mathbf{h}_{k,j}\sim \mathcal{CN}(0,\mathbf{R}_{k,j})$ , $\mathbf{R}_{k,j} \propto \mathbf{I}$. In the following, we present the results for the two services of web browsing and video separately.

\subsection{Web Browsing}
For web browsing service, we assume that the user number 1 to 10 are respectively receiving web page sizes of $18, 30, 50, 100, 200, 320, 400, 500, 650$ and $1000$ [kB]. Assume the minimum and maximum spectral efficiency for each user are $2$ [bit/s/Hz] and $7$ [bit/s/Hz], respectively. $K_1$ and $K_2$ in (\ref{web}) are obtained by assigning the minimum MOS to $R_{min}$ and the maximum to $R_{max}$ that results in $K_1=3.194$ and $K_2=15.1978$ (here we use the average $FS$= $320$ [kB] \cite{r16}). In addition, the \textit{MSS} and \textit{RTT} are $1460$ [byte] and $30$ [ms], respectively \cite{r16}. 

\begin{figure}
\centering
\includegraphics[width=0.5\textwidth]{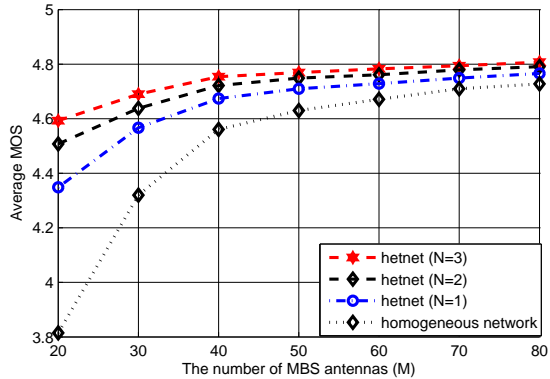}
\caption{Average MOS of the users vs. the number of the MBS antennas $M$ in a HetNet with $N_s=4$ SBS in comparison with a homogeneous network for web browsing service.}
\label{mergMOS}
\end{figure}

Fig. \ref{mergMOS} shows the average MOS of the users versus the number of the MBS antennas in a HetNet and compares it with a homogeneous network. The HetNet consists of one MBS and four small cells with different number of antennas. It should be noted that the results for homogeneous network are obtained by setting $N_s=0$. Here, to obtain the average MOS, the aggregated MOS of all users is divided by \textit{K} (total number of users). In this figure, we see that the average MOS of the HetNet is always better than the homogeneous network.
For example, in the case of \textit{M=20}, it is shown that adding small cells to the network leads to about 14\%-21\% improvement in the average MOS. In addition, increasing the number of MBS antennas from 20 to 80, improves the average MOS of the homogeneous network and HetNet($N=1$) for about 23\% and 9\%, respectively. In other words, these results show that employing small cells or MaMIMO MBS in the network leads to more user satisfaction. Also the need for adding small cells is much smaller when the BS has more antennas.

This figure also shows by increasing the number of the SBSs' antennas, the average MOS of the network will be improved. An interesting result that can be observed from this figure is that a good average MOS can be obtained either by a homogeneous MaMIMO MBS or a HetNet MaMIMO with less number of antennas at the MBS. For example, the average MOS of 4.5 is reached either by $M=40$ in homogeneous network or $M=20$ in a HetNet with four two-antenna SBSs.  

\begin{figure*}[t!]
\begin{subfigure}{.5\textwidth}
\centering
\includegraphics[width=1\textwidth]{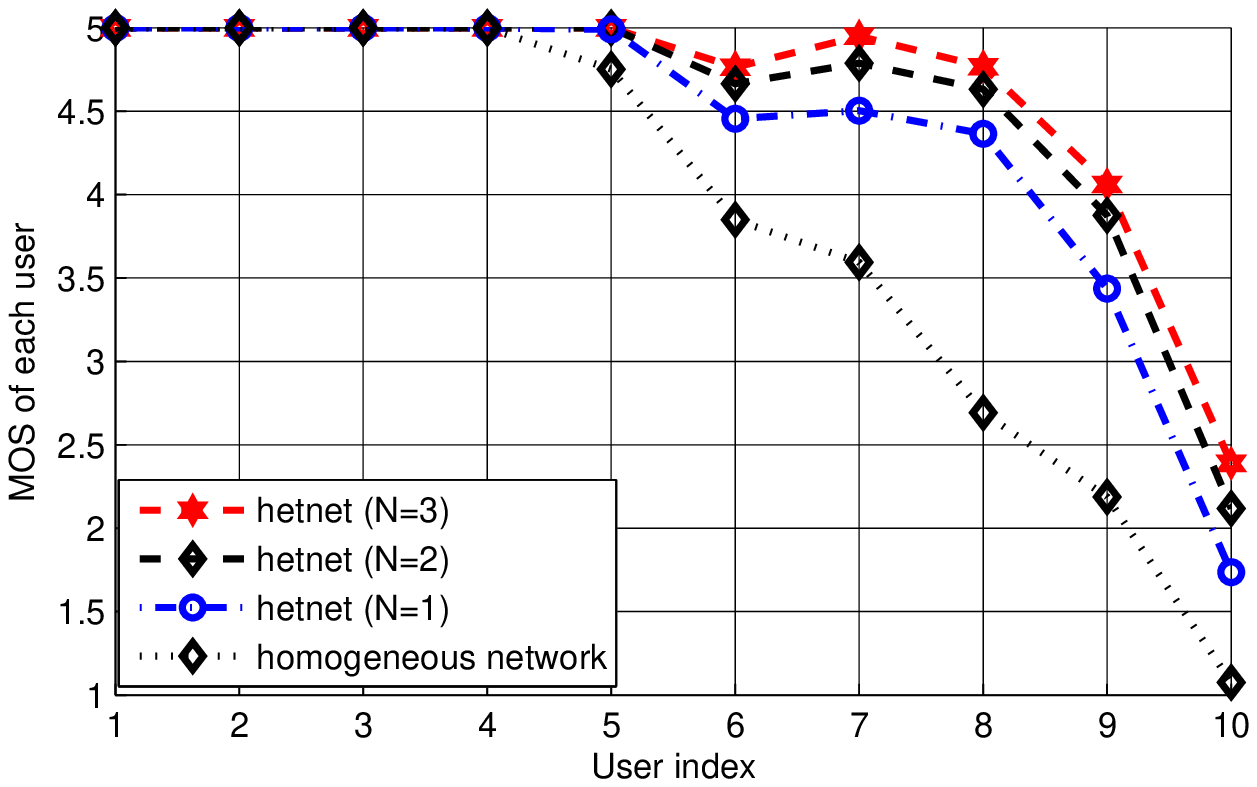}
\caption{$\mathrm{MOS}^{web}_{min,k}=1,\forall k$.}
\label{mos_user_20}
\end{subfigure}
\begin{subfigure}{.5\textwidth}
\centering
\includegraphics[width=1\textwidth]{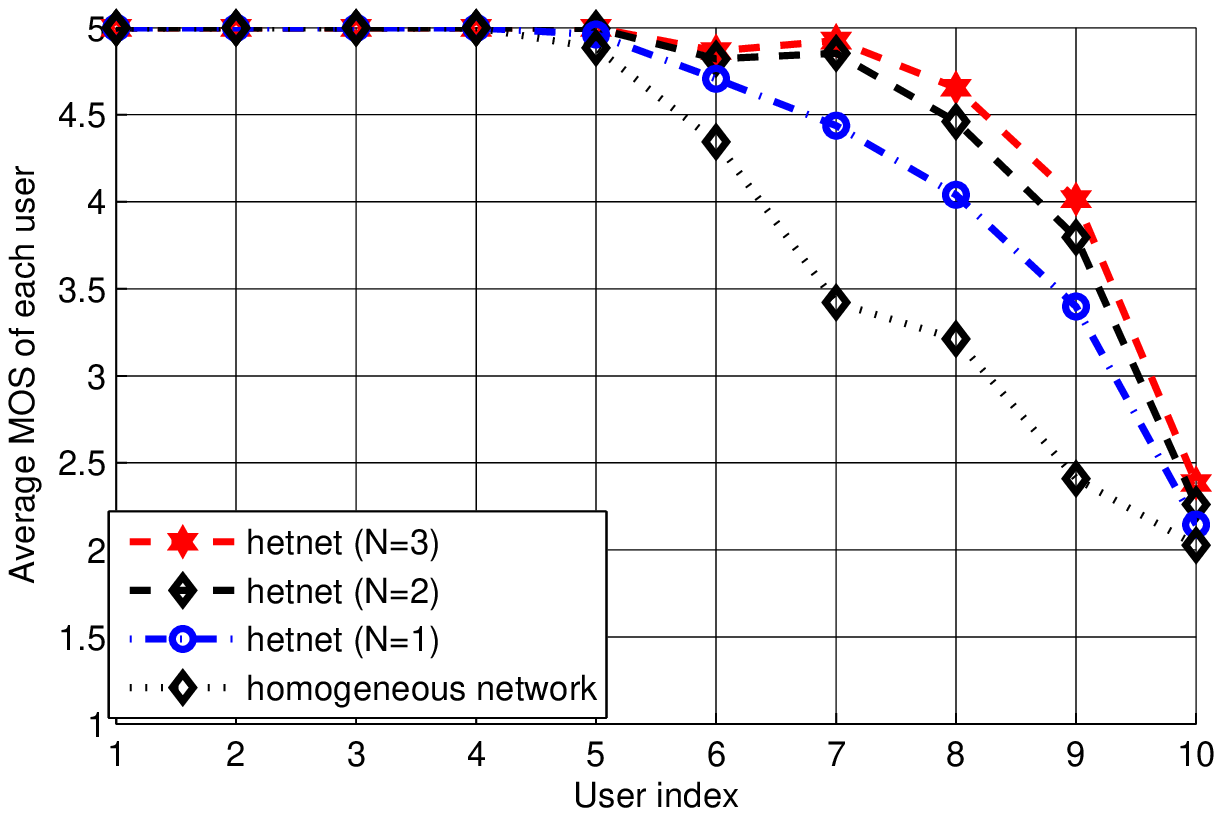}
\caption{$\mathrm{MOS}^{web}_{min,k}=2,\forall k$.}
\label{mos_user_20_alg2}
\end{subfigure}
\caption{\small MOS of the users in a network with $M=20$ antennas, $K=10$ users and different number of antennas at the SBSs for Web browsing service. The last four users are deployed in the small cells.}
\label{mos_user_20two}
\end{figure*}

Fig. \ref{mos_user_20} and \ref{mos_user_20_alg2} shows the performance of the proposed algorithm when $\mathrm{MOS}^{web}_{min,k}=1$ and $\mathrm{MOS}^{web}_{min,k}=2$, respectively. It should be noted that $\mathrm{MOS}^{web}_{min,k}=1$ implies that there is no QoE constraint in the optimization problems. In addition, the figures depict the MOS of each user in the MaMIMO HetNet where the MBS is equipped with 20 antennas, and the SBSs are equipped with 1, 2, and 3 antennas. Among the 10 users that are in the network, the first six users are always within the Macro cell coverage area and each of the last four users is within one small cell. Comparing Fig. \ref{mos_user_20} with Fig. \ref{mos_user_20_alg2} clarifies that the network with $\mathrm{MOS}^{web}_{min,k}=2$ is more robust to the sizes of web pages.
In other words, by considering QoE constraints in Fig. \ref{mos_user_20_alg2}, the MOS of the users are improved in comparison with the Fig. \ref{mos_user_20} (i.e. the case that there is not any QoE constraints). For example the user 10 with $FS=1000$ [kB] gets MOS of 1 and 2 in Fig. \ref{mos_user_20} and Fig. \ref{mos_user_20_alg2}, respectively.

\subsection{Video services}
In this part, we present the numerical results for video service. The network parameters are similar to the previous case. 
The parameters in (\ref{videowww}) and (\ref{psnrrrr}) are designed by using \cite{videoconstants} for PSNR between 30-42 dB. These parameters are $u=28.046$, $v=0.038$ and $r=5.024$. The parameters $g$ and $e$ in (\ref{videowww}) are also set to $27.37$ and $-39.43$, respectively.

\begin{figure}
\centering
\includegraphics[width=0.5\textwidth]{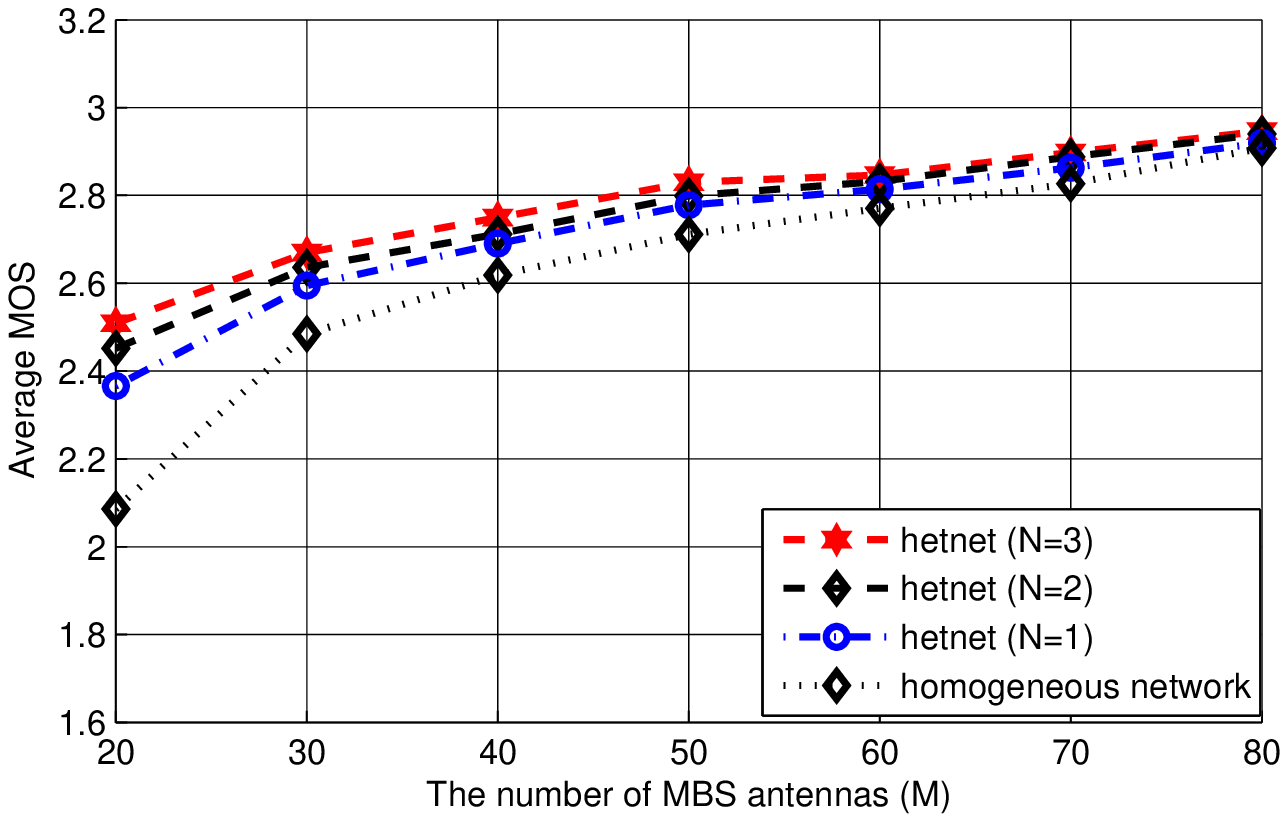}
\caption{Average MOS of the users vs. the number of the MBS antennas $M$ in a HetNet with $N_s=4$ SBS in comparision with a homogeneous network for video service.}
\label{totalMOS}
\end{figure}

Fig. \ref{totalMOS} shows the average MOS by considering $\mathrm{MOS}^{video}_{min,k}=1$ (i.e. without considering any QoE constraints in (\ref{videow2})). This figure compares the average MOS of a homogeneous network consisting of one MBS with a HetNet consisting of one macro cell and four small cells with different number of antennas. This figure indicates that for a given number of MBS antennas $M$, the average MOS of a HetNet is always better than in a homogeneous network. For example, for the case when $M=20$, it is shown that adding small cells to the network leads to about 13\%-20\% improvement in average MOS. In addition, increasing the number of MBS antennas from 20 to 80, improves the average MOS of the homogeneous network and HetNet (about 33\% and 19\%, respectively). In other words, these results show that employing small cells or MaMIMO MBS in the network leads to higher user satisfaction. Also the need for adding small cells is much smaller when the BS has more antennas.
This figure also shows by increasing the number of the antennas at the SBSs, the average MOS of the network will be improved. An interesting result that can be observed from this figure is that a good average MOS can be obtained either by a homogeneous MaMIMO MBS or a HetNet MaMIMO with less number of antennas at the MBS. For example, the average MOS of 2.71 is reached either by $M=50$ in homogeneous network or $M=40$ in a HetNet with four two-antenna SBSs. 

\begin{figure*}[t!]
\begin{subfigure}{.5\textwidth}
\centering
\includegraphics[width=1\textwidth]{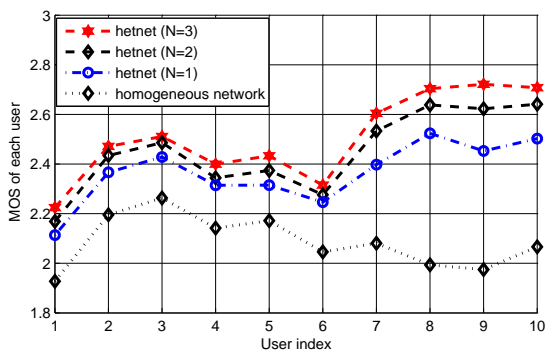}
\caption{$\mathrm{MOS}^{video}_{min,k}=1$.}
\label{usermos}
\end{subfigure}
\begin{subfigure}{.5\textwidth}
\centering
\includegraphics[width=1\textwidth]{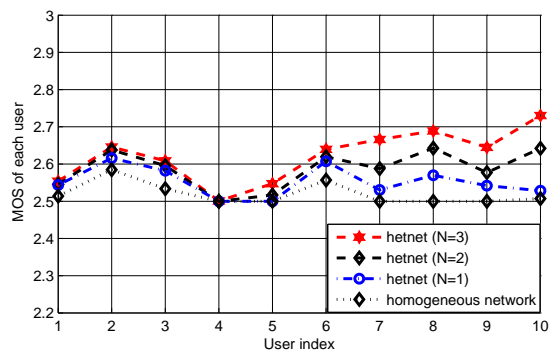}
\caption{ $\mathrm{MOS}^{video}_{min,k}=2.5$.}
\label{MOSusers2025}
\end{subfigure}
\caption{\small{MOS of the users in a network with $M=20$ antennas, $K=10$ users and different number of antennas at the SBSs for video service. The last four users are deployed in the small cells.}}
\label{video}
\end{figure*}

 Figs. \ref{usermos} and \ref{MOSusers2025} shows the performance of the proposed algorithm when $\mathrm{MOS}^{web}_{min,k}=1,\forall k$ and $\mathrm{MOS}^{web}_{min,k}=2.5,\forall k$, respectively. In addition, the figures depict the MOS of each user in the MaMIMO HetNet where the MBS is equipped with 20 antennas, and the SBSs are equipped with 1, 2, and 3 antennas. There are 10 users in the network which are numbered from 1 to 10 (on the horizontal axis). In addition, the first six users are always placed in the Macro cell coverage area and each of the last four users is within one small cell. These figures show the advantage of HetNets in improving MOS of each user which are concluded from solving the proposed scheme. 
In other terms, it can be seen in Fig. \ref{MOSusers2025} the MOS of the users are improved by considering QoE constraints in comparison with Fig. \ref{usermos} (when we do not consider QoE constraints). For example the user 10 gets the MOS of 2.07 and 2.51 in Fig. \ref{usermos} and \ref{MOSusers2025}, respectively (about 21\% improvement in QoE). Therefore the user 10 is more satisfied than before. 

\section{Conclusion}\label{sec:conclusion}
This paper proposed a joint beamforming and power allocation scheme for MaMIMO HetNets. We provided algorithms to optimize the beamfoming vectors at the MBS and SBSs based on maximizing the aggregated MOS of the network. We consider two different services of web browsing and video in the network.
 The resulting optimization problems of two provided algorithms were not convex, hence we transformed them to convex optimization problem to design the beamforming vectors. Our simulation results show that increasing the number of antennas at the MBS or SBSs leads to a better QoE of users. In addition, we showed that the QoE can be improved by adding small cells to the homogeneous network in both web browsing and video services. 
 
\begin{thebibliography}{100}
\bibitem{r3}
I. Hwang, B. Song, and S. S. Soliman,
\newblock ``A holistic view on hyper-dense heterogeneous and small cell networks,''
\newblock {\em IEEE Communications Magazine}, vol. 51, no. 6, pp. 20-27, June 2013.
\bibitem{r1-new}
M. Agiwal, A. Roy, and N. Saxena,
\newblock ``Next generation 5G wireless networks:A comprehensive survey,''
\newblock  {\em IEEE Communications Surveys and Tutorials}, vol. PP, no. 99, pp. 1-40, Feb. 2016.
\bibitem{TVT-traffic}
Q. Wu, Z. Du, P. Yang, Y. D. Yao, and J. Wang, 
\newblock ``Traffic-Aware online network selection in heterogeneous wireless networks,''
\newblock {\em IEEE Transactions on Vehicular Technology}, vol. 65, no. 1, pp. 381-397, Jan. 2015.
\bibitem{TVT-qoe-aware}
D. Wu, Q. Wu, Y. Xu, and Y. C. Liang,
\newblock ``QoE and Energy aware resource allocation in small cell networks with power selection, load management and channel allocation,''
\newblock {\em IEEE Transactions on Vehicular Technology}, vol. 66, no. 8, pp. 7461-7473, Jan. 2017.
\bibitem{TVT-adhoc}
P. T. Quang, K. Piamrat, K. D. Singh, and C. Viho,
\newblock ``Video streaming over Ad-hoc networks: a QoE-based Optimal Routing Solution,''
\newblock {\em IEEE Transactions on Vehicular Technology}, vol. 62, no. 2, pp. 1533-1546, Apr.  2016.
\bibitem{r8-new}
J. Shaikh, M. Fiedler, and D. Collange,
\newblock ``Quality of Experience from user and network perspectives,''
\newblock {\em Annals of Telecommunications}, vol. 65, no. 1, pp. 47-57, Feb. 2010. 
\bibitem{r9-new}
P. Ameigeiras, J. J. Ramos-Munoz, J. Navarro-Ortiz, P. Mogensen, and M. Lopez-Soler,
\newblock ``QoE oriented cross-layer design of a resource allocation algorithm in beyond 3G systems,''
\newblock {\em Computer Communications}, vol. 33, no. 5, pp. 571-582, Mar. 2010. 
\bibitem{r16}
M. Rugelj, U. Sedlar, M. Volk, J. Sterle, M. Hajdinjak, and A. Kos,
\newblock ``Novel cross-layer QoE-aware radio resource allocation algorithms in multiuser OFDMA systems,''
\newblock  {\em IEEE Transactions on Communications},  vol. 62, no. 9, pp. 3196-3208, Sep. 2014.
\bibitem{r12}
A. Kazerouni, F. J. Lopez-Martinez, and A. Goldsmith,
\newblock ``Increasing capacity in Massive MIMO cellular networks via small cells,''
\newblock  in {\em Proc. Global Communications Conference Workshops.}, 2014, pp. 258-363.
\bibitem{r14}
E. Bj\"{o}rnson, M. Kountouris, and M. Debbah,
\newblock ``Massive MIMO and small Cells: improving energy efficiency by optimal soft-cell coordination,''
\newblock  in {\em Proc.  International Conference on Telecommunications (ICT)}, 2013, pp. 1-5.
\bibitem{TWC1}
Z. Du, Q. W,u, P. Yang, Y. Xu, J. Wang, and Y. D. Yao,
\newblock ``Exploiting user demand diversity in heterogeneous wireless networks,''
\newblock  {\em IEEE Transactions on Wireless Communications},  vol. 14, no. 8, pp. 4142-4155, Aug. 2015.
\bibitem{relay}
S. Lin, W. Sheen, and C. Huang,
\newblock ``Downlink performance and optimization of relay-assisted cellular networks,''
\newblock in {\em Proc. Wireless Communications and Networking Conference (WCNC)}, 2009, pp. 5-8.
\bibitem{int16}
S. Thakolsri, W. Keller, and E. Steinbach, 
\newblock ``Application-driven cross layer optimization for wireless networks using MOS-based utility functions,''
\newblock  in {\em Proc. Communications and Networking in China (ChinaCOM.)}, 2009, pp. 1-5.
\bibitem{r17}
Y. Deyu, S. Mei, T. Yinglei, W. Xiaojun, and L. Guofeng,
\newblock ``QoE-oriented resource allocation for ,multiuser- multiservice femtocell networks,''
\newblock  {\em China Communications},   vol. 12, no. 10, pp. 27-41, Oct. 2015.
\bibitem{int23}
K. Wu, L. Guo, T. Song, and J. Lin, 
\newblock ``Bio-Inspired multi-user beamforming for QoE provisioning in cognitive radio networks,''
\newblock  in {\em Proc. }Network Infrastructure and Digital Content (IC-NIDC) International conference, 2012, pp. 173-177.
\bibitem{Twc2}
R. Imran, M. Odeh, N. Zorba, and C. Verikoukis,
\newblock ``Quality of Experience for spatial cognitive systems within multiple antenna scenarios,''
\newblock {\em Transactions on Wireless Communications}, vol. 12, no. 8, pp. 4153-4161, Aug. 2013.
\bibitem{hetero}
D. Wu, Q. Wu, Y. Xu, J. Jing, and Z. Qin,
\newblock ``QoE-Based Distributed Multichannel Allocation in 5G Heterogeneous Cellular Networks: A Matching-Coalitional Game Solution,''
\newblock  {\em IEEE Access},  vol. 5, pp. 61-71, Sep. 2016.
\bibitem{videoconstants}
L. U. Choi, M. T. Ivrlac, E. Steinbach, and J. A. Nossek,
\newblock ``Sequence-level models for distortion-rate behaviour of compressed video,''
\newblock  in {\em Proc.  IEEE International Conference on Image Processing}, 2005, pp. II - 486-9.
\bibitem{perantenna}
S. Zhang, R. Zhang, and T. J. Lim,
\newblock ``Massive MIMO with per-antenna power constraint,''
\newblock in {\em Proc. Global Conference on Signal and Information Processing (GlobalSIP)}, 2014, pp. 642-646.
\bibitem{rankw}
E. Bj\"{o}rnson, N. Jalden, M. Bengtsson, and B. Ottersten, “Optimality
\newblock ``Optimality properties, distributed strategies, and measurement-based evaluation of coordinated multicell OFDMA transmission,''
\newblock  {\em  IEEE Transactions of Signal Processing},  vol. 59, no. 12, pp. 6086-6101, Jan. 2011.
\bibitem{r19}
T. Lv, H. Gao, R. Cao, and J. Zhou,
\newblock ``Coordinated secure beamforming in K-user interference channel with multiple eavesdroppers,''
\newblock  {\em  IEEE Communications Letters},  vol. 5, no. 2, pp. 212–215, Jan. 2016.
\bibitem{converge}
A. Beck, A. Ben-Tal, and L. Tetruashvili,
\newblock ``A sequential parametric convex approximation method with applications to nonconvex truss topology design problems,''
\newblock  {\em Journal of Global Optimization}, vol. 47, no. 1, pp. 29-51, May, 2010.
\end{thebibliography}
\end{document}